\documentclass[runningheads,a4paper]{llncs}

\usepackage{amssymb}
\usepackage{graphicx}
\usepackage{subfigure}
\usepackage{hyperref}

\usepackage{url}
\urldef{\mailsa}\path|{ruiguo, wangzh, zlc, lijzh, honggao}@hit.edu.cn|
\newcommand{\keywords}[1]{\par\addvspace\baselineskip
\noindent\keywordname\enspace\ignorespaces#1}

\begin{document}

\mainmatter  % start of an individual contribution


\title{Harbinger:\\An Analyzing and Predicting System for Online Social Network Users' Behavior}

%\footnote{need to change?}
\titlerunning{\ }

\author{Rui Guo\and Hongzhi Wang\and Lucheng Zhong\and Jianzhong Li\and Hong Gao}
\authorrunning{\ }

\institute{Harbin Institute of Technology\\
Harbin, Heilongjiang, China\\
\mailsa}
%\url{http://www.springer.com/lncs}}


\toctitle{Lecture Notes in Computer Science}
\tocauthor{Authors' Instructions}
\maketitle

%\begin{abstract}
%The abstract should summarize the contents of the paper and should
%contain at least 70 and at most 150 words. It should be written using the
%\emph{abstract} environment.
%\keywords{We would like to encourage you to list your keywords within
%the abstract section}
%\end{abstract}

\begin{abstract}
Online Social Network (OSN) is one of the hottest innovations in the past years, and the active users are more than a billion. For OSN, users' behavior is one of the important factors to study.
This demonstration proposal presents \textit{Harbinger}, an analyzing and predicting system for OSN users' behavior.
In \textit{Harbinger}, we focus on tweets' timestamps (when users post or share messages), visualize users' post behavior as well as message retweet number and build adjustable models to predict users' behavior.
Predictions of users' behavior can be performed with the discovered behavior models and the results can be applied to many applications such as tweet crawler and advertisement.

\end{abstract}

\section{Introduction}
Online social networks have exploded incredibly in the past years.
For instance, Twitter has 200 million active users who post an average of 400 million tweets every day~\cite{twitterCite}.
%and Facebook reports 1.11 billion monthly active users as of March 2013~\cite{facebookCite}.
Since the large group of users make OSNs valuable for both commercial and academical applications, the understanding of users' behaviors could help these applications to improve efficiency and effectiveness.
%For example, advertisement and user information collection, require to analyze target users' behavior to increase efficiency by finding the chances when many users are online or post new messages.

The understanding of users' behaviors brings challenges.
The crucial one is that they keep on challenging computing resource.
For billions of active OSN users, various models or parameters are required to describe users' different behaviors. Another difficulty is that users are influenced by many factors, most of which are invisible through OSNs. It makes users' behavior difficult to predict.
%including work and entertainment time, user's mood and even weather.

Existing works study OSN users' behavior in several different ways. \cite{charactering1} characterizes behavior by clickstream data.
They summarized HTTP sessions from an aggregation website. Its conclusion is that browsing counts 92\% of all users' activities, but this observation cannot be obtained from public OSN data.
\cite{measuring2} downloaded user profile pages, and modeled users' online time with Weibull distributions.
%\cite{user3} analyzed user interaction graphs in Facebook, and prove ``small-world'' properties in social graph counterparts: users tend to interact with small subset of friends more frequent, often no interactions with up to 50\% of their Facebook friends.
%\cite{zaman2010predicting} studies retweet number and time, and build 3 different models to predict the future retweet number. However, both the tweeter, a retweeter, and the content of the tweet are required to be input, which is difficult to collect in the real OSNs. And the output is a value $p$ as the probability of a retweet of the tweet by the retweeter. The model is costly and inefficient.

%costs a lot of computing resource to calculate the possible retweet number.
%\fn{what is the shortcomings of current systems}
%%%%%%%%%%%%%%%%%下面这段是新加的，关于本demo的必要性
To analyze and predict users' behavior, we present \textit{Harbinger} system. \textit{Harbinger} has two major functions: users' post behavior analyzing and single message retweet number analyzing.
Statistics methods in~\cite{rui2013cuvim} are applied to avoid the effect of invisible factors.
We observe both a group of users and single tweets, collect message timestamps and message retweet numbers, visualize users' post behavior
and the variation of message retweet number, and describe them by Gaussian Mixture Model and Logarithm Model.
Unlike previous works, \textit{Harbinger} analyzes users' post behavior through tweets' timestamps rather than users' clickstream, online time or friendship.

The remainder of the demonstration proposal is organized as follows. In Section 2, we overview \textit{Harbinger} and introduce system architecture. In Section 3, we present mathematics models and algorithms. In Section 4, we give demonstration scenario of the system.

%For users, we study the frequency of new messages in a specific period, build, and apply to get the and predict user behavior in the future.
%For a single tweet, we crawled it once ten minutes to collect the retweet number at each time point, and develop an logarithm function of time to represent retweet number and influence of this tweet.

\section{System Overview}

%\subsection{Basic Architecture}
Our System contains two major functions: the users' post behavior analyzing function and the single message retweet number analyzing function.

For the OSN users' post behavior, the user of \textit{Harbinger} is expected to choose an analyzing function, a target OSN user and an analyzing pattern (daily, weekly or monthly pattern). Then corresponding data are selected and statistics are preformed. We set the statistics time span to be one hour (from 00:00 to 24:00) to the daily pattern, and one day from the beginning to the end of the week or month to the weekly or monthly pattern. Finally, the figure of message number and time, and the analysis results such as Figure~\ref{fig:a} will be plotted.

For the retweet number function, a user chooses a tweet function and target tweet rather. The system analyzes the data and plots the relationship between time and the retweets of the selected tweet such as Figure~\ref{fig:b}.

% And Logarithm Model will be made in the analyze module rather GMM as well.

\begin{figure}
  \centering
  \includegraphics[width=0.3\textwidth]{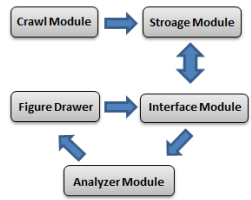}
  \caption{Modules of \textit{Harbinger}}
  \label{fig:mod}
\end{figure}

As shown in Figure~\ref{fig:mod}, \textit{Harbinger} has five major modules: crawl, storage, interface, analyzer and figure drawer modules.

In the crawl module, we develop an OSN crawler through OSN official API to collect information and this is also an application for our prediction model.
The crawler collects the message information, including content and timestamp, and then stores it in the storage module, where all data are stored in a database.
The User Interface (UI) of the Interface Module connects visitors to \textit{Harbinger} and other modules.
The user of \textit{Harbinger} can select target OSN user or tweet, and the analyzing pattern in the UI module. The selection is sent to the analyzer.
After statistics, analysis and calculating in the analyzer are based on the models in Section~\ref{sec:model}, and the results are sent to the figure drawer, which draws figures in the UI according to the analysis results.  %GMM and EM will also be conducted in the analyze module if visitor chooses the daily pattern.

\section{Models and Algorithms}
\label{sec:model}
\begin{figure}
    \centering
    \subfigure[Figure of Gaussian Mixture Model]{
        \label{fig:a}
        \includegraphics[width=0.47\textwidth]{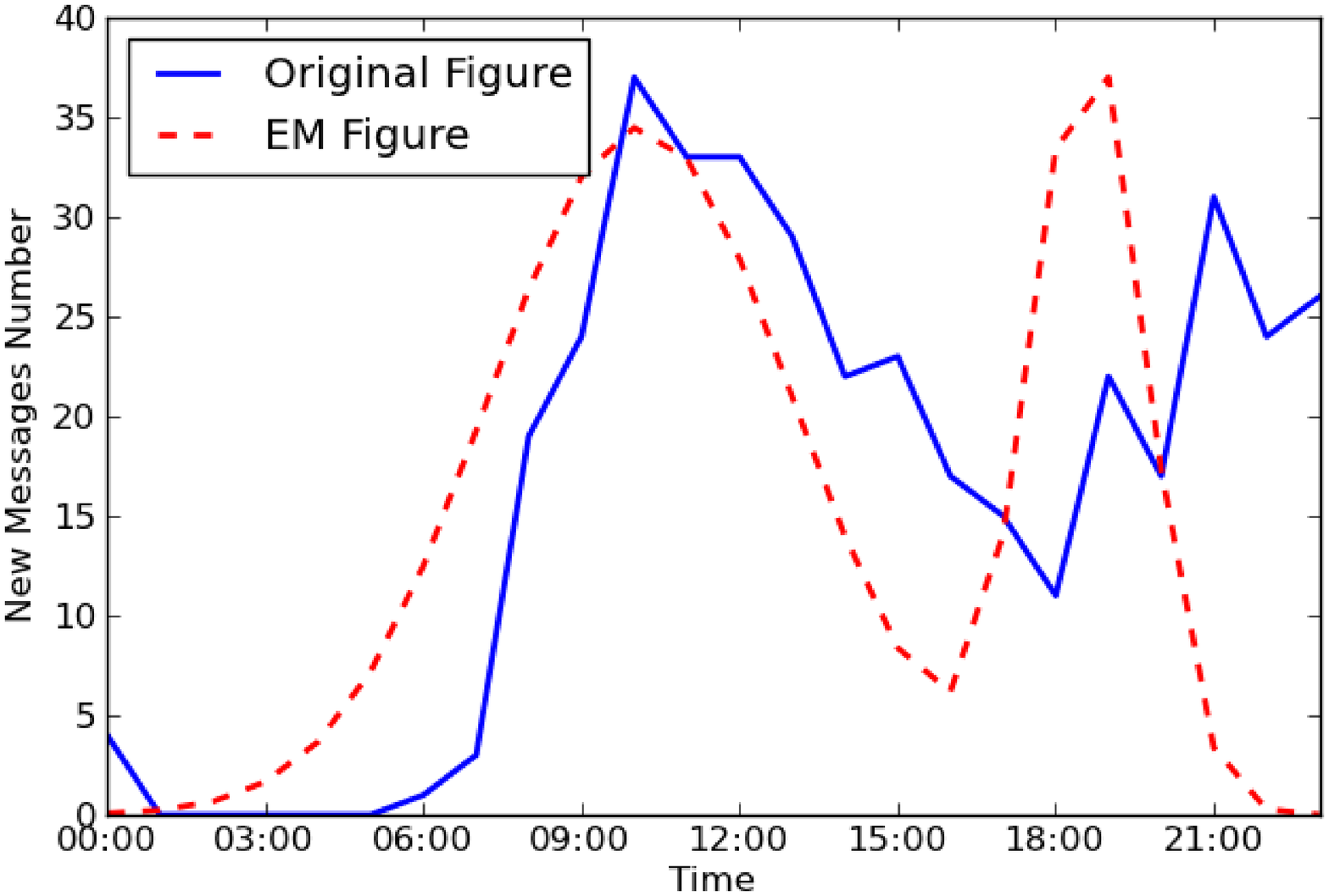}
    }
    \hspace{\fill}
    \subfigure[Figure of Logarithm Model]{
        \label{fig:b}
        \includegraphics[width=0.47\textwidth]{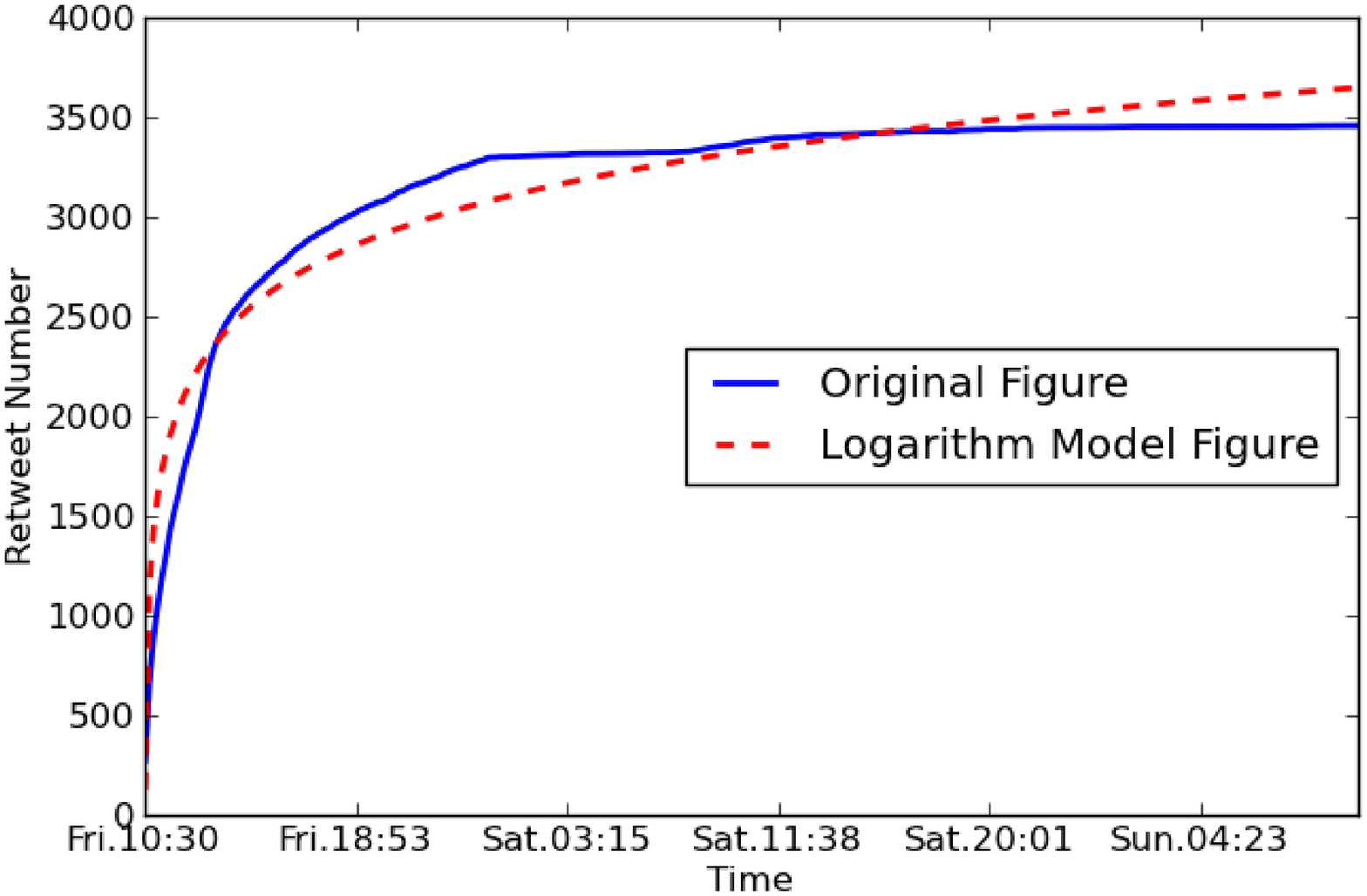}
    }
    \caption{Figure of retweet number and time}
    \label{fig:ab}
\end{figure}

In this section, we describe the models and algorithms used in our system. They are the major parts of the analyzer. Based on the technology in~\cite{rui2013cuvim}, we develop the Gaussian Mixture Model (GMM) to describe OSN user behavior and Logarithm Model to illustrate the relationship between retweet number and time.
%
%\footnote{add a sentence}
Figure~\ref{fig:ab} shows the results of the GMM and Logarithm Model.

\subsection*{Gaussian Mixture Model }
In the daily pattern of the users' post behavior, we find that the relationship between new messages' number and the time in a day follows the addition of two Gaussian Distributions (or Normal Distributions).
OSN users often work during the day, and rest at noon and dusk. Thus there are two peaks of fresh OSN messages. The curve around each peak is similar to a Gaussian Distribution.
%As an example, it can be explained that the closer to entertainment time, the more frequent the user visits OSN and post messages.
As the result, the figure can be treated as a mixture of two Gaussian Distributions.

Thus we develop Gaussian Mixture Model~\cite{wikiGMM} (GMM) to compute unknown parameters of the figure.
%GMM is a mature method to deal with unknown parameters, and it is well applied in machine learning, nature language proceeding and computer graphic.
Assume the daily time is $t$, the number of new messages during $t$ is $f(t)$, the two Gaussian Distributions are $N_1(\mu_1,\sigma^{2}_1 )$ and $N_2(\mu_2,\sigma^{2}_2 )$, and there is
\begin{displaymath}
f(t)=\frac{1}{\sqrt{2\pi}\sigma_1}e^{-\frac{(x-\mu_1)^2}{2\sigma_1^2}}+\frac{1}{\sqrt{2\pi}\sigma_2}e^{-\frac{(x-\mu_2)^2}{2\sigma_2^2}}
\end{displaymath}

To figure out the exact parameters ($\mu_1,\sigma_1, \mu_2,\sigma_2 $) in GMM, we apply Expectation-Maximization (EM) algorithm~\cite{wikiEM}, which is the computing process of GMM.
%\footnote{add a sentence}
Figure~\ref{fig:a} shows the results of GMM. In Figure~\ref{fig:a}, the solid line means the original figure of time and retweet number (the user post frequency in a day), and the dotted line means the results of EM algorithm. The results show that the post frequency is indeed similar to the sum of two Gaussian distributions.

%Assume a set $X$ of observed data, a set $Z$ of unobserved latent data, a vector of unknown parameters $\theta$, and a likelihood function $L(\theta;X,Z)=p(X,Z|\theta)$, the maximum likelihood estimate (MLE) of the unknown parameters is determined by the marginal likelihood of the observed data:
%\begin{displaymath}
%L(\theta;X,Z)=p(X|\theta)=\sum_{Z}{} p(X,Z|\theta)
%\end{displaymath}

%The EM algorithm aims to find those parameters iteratively according to the following two steps:
%
%\textit{Expectation step (E step)} calculates the expected value of the log likelihood function, with respect to the conditional distribution of under the current estimate of the parameters.
%
%\textit{Maximization step (M step)} finds the parameter that maximizes this quantity.
%
%The EM algorithm halts in a given rounds of iteration or a given precision.
%

\subsection*{Logarithm Model}
We find that the relation between retweet number and posted time of a specific message follows the logarithm function. After stretching and shifting, a basic logarithm function can describe retweet number properly. In this curve, x-axis is the posted time (e.g. how long the tweet is posted) of the message and y-axis is the retweet number.

The retweet number grows very fast after the tweet is posted, and with x increases, the growing becomes more and more unchanged.
Thus we compute the retweet number $RN$ in the posted time $t$ (e.g. how long the message is posted) as $RN = k_1 log_{base} (k_0 x + k_2) + k_3$,
where $base$ is the base of logarithm function representing the steepness of the curve. $k_0$ is the x-axis stretch parameter, $k_1$ is the y-axis stretch parameter, $k_2$ is the x-axis shift parameter, and $k_3$ is the y-axis shift parameter.

To compute the exact parameters of Logarithm Model, we apply Least Squares Algorithm~\cite{wikiLS}. The basic idea of Least Squares algorithm is to approximate the model by a linear function and to refine the parameters with iterations.

%\footnote{add a sentence}
Figure~\ref{fig:b} describes the result of Logarithm Model. The solid line means the original figure of time and retweet number growth (how the retweet number changes over time after being posted), and the dotted line means the result of Logarithm Model. The result shows that the original figure is indeed similar to a logarithm function figure.

\section{Demonstration Scenario}
\textit{Harbinger} is well encapsulated with a friendly interface. Though the system has a large database and specialized analyzer module, what the user faces is only a simple UI.
%\textit{Harbinger} just requires users to click buttons rather than input declarative query languages.
The interface is shown in the video (http://www.youtube.com/watch?v=x\\gXcsbNYqqQ\&feature=youtu.be).
%We aims to make \textit{Harbinger} both simple and practical.

The post behavior analyzing function is selected with the radio button.
The user of \textit{Harbinger} can select ``All User'' or a specific user from the target OSN user list to study all OSN users' behavior or a specific user's behavior.
%Then all the tweets posted by the target OSN user are listed in the tweet list.
And the figure of the distribution is plotted at the bottom of the interface according to the parameters shown at the right side. To show the single message retweet number analyzing function, the other radio button is selected as well as a target message.

\bibliographystyle{abbrv}
\bibliography{main}
\end{document}